\documentclass[a4paper,11pt,reqno]{amsart}

\numberwithin{equation}{section}

\usepackage[]{amsmath}
\usepackage{amsfonts}
\usepackage{amssymb}
\usepackage{xcolor}
\usepackage{mathrsfs}

\usepackage{tikz}
\usetikzlibrary{shapes.geometric,positioning,calc}

\usepackage{binarytree} 

\usepackage[margin=3cm]{geometry} 
\theoremstyle{definition}

\renewcommand{\i}{{\mathrm{i}}}

\def\Phibar{ \overline{\Phi} }
\def\Phitld{ \widetilde{\Phi} }
\def\Stld{ \widetilde{S} }
\def\Jtld{ \widetilde{J} }
\def\Htld{ \widetilde{H} }
\def\Ktld{ \widetilde{K} }
\def\Ztld{ \widetilde{Z} }
\def\oPsi{ \overline{\Psi} }
\def\oPhi{ \overline{\Phi} }
\def\bedt{    \beta\hskip-1.1ex\raisebox{1.8ex}{\scalebox{.5}{$\,\bullet\,$}}  }
\def\one{\hbox{{1}\kern-.25em\hbox{l}}}

\title[Algebraic inversion of the Duffin-Kemmer equation] 
 {Gauge invariant formulation of the self-interacting Duffin-Kemmer-Petiau equations.}
\author[Jarvis and Inglis]{P. D. Jarvis${}^*$ and S. M. Inglis}

\address{School of Natural Sciences,
(Mathematics \& Physics), University of Tasmania}
\email{peter.jarvis@utas.edu.au, Shaun.Inglis@utas.edu.au}

\date{\today \\${}^*$ Alexander von Humboldt Fellow}
\begin{document}

\begin{abstract}
We show that the Duffin-Kemmer-Petiau equation, minimally coupled to an Abelian gauge field, can be regarded as a matrix equation for the gauge potential produced internally from the matter fields. This can be solved as a rational expression in terms of currents bilinear in the matter wavefunction, together with a similar expression for the field strength tensor, thus providing a gauge invariant formulation of the self-interacting DKP equations. We give the derivation of this result for the 5 component DKP system, by analogy with the Dirac equation case. To this end, we establish the algebraic structure of the set of bilinear currents, and the properties of the minimal generating set, which consists of two scalars and two four-vectors, together with a single quadratic constraint.
\end{abstract}
\maketitle
\vfill

\pagebreak

\section{Introduction} 
The Dirac equation, minimally coupled with an electromagnetic field,
can be regarded as a set of algebraic equations for the gauge potential, whose solution is a rational expression in terms of currents bilinear in the Dirac wavefunction, 
and their derivatives. This result was obtained by Radford \cite{Radford1996}\,,
and subsequently developed in higher dimensional \cite{BoothLeggJarvis2001}\,
and non-Abelian cases \cite{InglisJarvis2012}. The resulting Maxwell-Dirac equations have been 
shown to admit monopole-like solutions \cite{Radford1996,BoothRadford1997,RadfordBooth1999, Radford2003,InglisJarvis2014}\,.

In this paper we provide the corresponding inversion construction for the self-interacting Duffin-Kemmer-Petiau (DKP) equation \cite{duffin1938characteristic,kemmer1939particle,
petiau1938university}. We concentrate here on the 5 component representation, with the 10 component system to be treated in a separate work. 

In section \ref{sec:DKPalg} below, we briefly review the DKP equation and the DKP algebra,
and study the algebra of linearly independent bilinear currents, and that of their
algebraically independent generating set, together with the 
Fierz-DKP \cite{Fierz1937} rearrangement identities appropriate to the 5 component system. In section \ref{sec:DKPinv} these results are used to obtain the 
expressions for the gauge potential and the field strength tensor, and hence
arrive at a gauge-invariant formulation of the self-interacting DKP equations.

\section{DKP equation and DKP algebra}
\label{sec:DKPalg}
The Dirac equation together with the DKP equation are the unique instances of relativistic first order equations based on wave functions belonging to representations of a five-dimensional orthogonal group \cite{LubanskiI1942,LubanskiII1942,bhabha1945relativistic} which describe single-mass systems. When interactions are introduced through
minimal coupling to an Abelian gauge potential, 
\begin{equation}
\big(i(\partial^\mu + ie A^\mu)\Gamma_\mu + m\big)\Phi =0\,,
\end{equation}
it is notable that there is a simple rearrangement whereby the system
can be viewed as a linear matrix equation for the potential,
$R^\mu A_\mu = \Psi$\,, for which a matrix inversion, if it exists, would yield
an algebraic expression $A_\mu = (R^{-1})_\mu \Psi$\, for the gauge potential itself.
Here $R^\mu \equiv \Gamma^\mu\Phi$ is the rectangular matrix of coefficients of the potential, and $\Psi$ represents the  terms independent of the potential, occurring in the equation.
As mentioned above, this procedure can indeed be implemented in the case of the
Dirac equation, and the solution for the gauge potential is a rational expression in terms of a set of real tensor quantities, or `current bilinears', which are quadratic in the Dirac wavefunction and its gradient. These currents are central to classical interpretations of the Dirac
equation in `relativistic fluid' formulations \cite{Takabayasi1957,halbwachs1960theorie}\,, and have been analyzed in this context by Crawford \cite{Crawford1985}\,.  

In practice, the inversion of the coefficient matrix in the Dirac case ($\Gamma_\mu\equiv \gamma_\mu $\,) proceeds indirectly, by using properties of 
the Dirac algebra or Clifford algebra of $\gamma_\mu$ matrices,
\begin{equation}
\gamma_\mu\gamma_\nu + \gamma_\nu\gamma_\mu=2 \eta_{\mu\nu}{\mathbb I}\,,
\end{equation} 
where $\eta_{\mu\nu}:=\mathrm{diag}(1,-1,-1,-1)$ is the flat spacetime Minkowski metric. In this paper we show that in the DKP case 
($\Gamma_\mu\equiv \beta_\mu $\,) analogous manipulations are also possible,
starting with the defining relations of the Kemmer $\beta_\mu$ matrices, namely
\begin{equation}
\beta_\mu\beta_\rho \beta_\nu + \beta_\nu\beta_\rho \beta_\mu 
= \eta_{\mu\rho}\beta_\nu +\eta_{\nu\rho}\beta_\mu\,.
\end{equation}
The algebraic structure of the $\beta_\mu$ was analyzed by Kemmer \cite{kemmer1939particle}\,, and in particular in great detail by Harish-Chandra \cite{harish1946correspondence}\,.
An immediate effect of the fact that the equations are not inhomogeneous, 
is the existence of a 1-dimensional representation with $\beta_\mu=0$\,,
and indeed \cite{kemmer1939particle,harish1946correspondence} the 126-dimensional enveloping algebra splits into $1-$\,, $25-$\, and 
$100-$ dimensional sectors spanned by the $1-$\,, $5-$\, and 
$10-$ component irreducible DKP representations. Below, we proceed with an investigation of the interacting DKP equation for the 5 component system, with the 10 component system to be treated in a later work.

Analyzing the Casimir operator eigenvalues of the Lorentz symmetry algebra generators
$\textstyle{\frac 14}{[} \beta_\mu, \beta_\nu {]}$\,, or adopting a concrete matrix basis,
reveals in particular that, for the 5 component case, the combination ${\mathbb I}-\beta^\mu\beta_\mu \equiv {\mathbb I}- \beta^2$ is a projector \cite{kemmer1939particle}; in consequence, any element in the DKP enveloping algebra can be resolved covariantly into block form, corresponding to mappings between its eigenspaces. Carrying this out for the generators $\beta_\mu$\, leads
to the definition of the companion generators
\begin{equation}
\bedt_\mu := \textstyle{\frac 13}\big(\beta_\mu \beta^2 - \beta^2 \beta_\mu\big)\,.
\end{equation}
The set $\{ {\mathbb I}, \beta_\mu, \beta_\mu\beta_\nu, \bedt_\mu, \beta^2 \}$\, (where 
$\beta^2 = \beta^\nu\beta_\nu$)\, generates a basis (of 25 linearly independent elements) of the algebra. These elements are not trace-orthogonal, as a consequence of the reducibility of 
$\beta_\mu\beta_\nu$, and we have
\begin{align}
Tr\big(\beta_\mu\beta_\nu\big) = &\, 2\eta_{\mu\nu}=-Tr\big(\bedt_\mu\bedt_\nu\big)\,;\\
Tr\big(\beta_\kappa\beta_\lambda\beta_\mu\beta_\nu\big) =&\,
\eta_{\kappa\lambda}\eta_{\mu\nu} + \eta_{\kappa\mu}\eta_{\lambda\nu}\,,
\end{align}
and others zero to this degree. Using these trace identities, and the DKP algebra defining relations, allows elements of the DKP algebra, of any degree in the $\beta_\mu$ and
$\bedt_\mu$\,, to be re-written in terms of the basic set. A compilation of such identities is given in the Appendix.

Finally, following Kemmer \cite{kemmer1939particle}, we introduce the real, symmetric, involutive matrix $\eta$ which implements the equivalence of $\beta_\mu$ with its transpose, for which
\begin{equation}
\label{eq:EtaTransposeEq}
\beta_\mu = \eta \beta^\top_\mu \eta\,; \quad \bedt_\mu = -\eta \bedt{}^\top_\mu \eta\,,
\end{equation}
we define the charge conjugate wavefunction $\Phibar := \Phi^\dagger \eta$\,, and introduce the set of real bilinear DKP currents: scalars $S$\,, $S^\flat$\,; charge vector current $J_\mu$\, and companion vector current  $H_\mu$\,; and tensor current $K_{\mu\nu}$\, (with $H^{*}_{\mu}=-H_{\mu}$ and $K^{*}_{\mu\nu}=K_{\nu\mu}$), defined as follows:
\begin{equation}
S:= \Phibar\Phi\,,\quad S^\flat := \Phibar\beta^2 \Phi\,,\quad
J_\mu := \Phibar\beta_\mu\Phi\,,\quad H_\mu := \Phibar\bedt_\mu\Phi\,,\quad
K_{\mu\nu} := \Phibar\beta_\mu\beta_\nu\Phi\,,
\end{equation}
(with $\eta^{\mu\nu}K_{\mu\nu} \equiv  S^\flat $)\,. Correspondingly, from the
trace properties, we extract the Fierz-DKP rearrangement identity 
\begin{align}
\label{eq:FDKPidentity}
\Phi\Phibar = &\,
\big( \textstyle{\frac 59} S - \textstyle{\frac 29}S^\flat\big) {\mathbb I} 
+\textstyle{\frac 12}J^{\mu}  \beta_\mu + K^{\nu\mu} \beta_\mu\beta_\nu
 - \textstyle{\frac 12}H^{\mu}  \bedt_\mu 
 - \big(\textstyle{\frac 29}S +\textstyle{\frac 19}S^\flat\big) \beta^2 \,.
\end{align}
(see Appendix for details). Using this identity, the expansion of products of the form $(\Phibar \Delta \Phi)\cdot (\Phibar \Delta' \Phi)$\,,
where $\Delta\,, \Delta'$ are DKP matrices, generates a system of homogeneous quadratic relations, or Fierz-DKP identities, expressing the algebraic dependence 
amongst the current bilinears. For example, if $\Delta=\Delta'= {\mathbb I}$\,,
we have immediately
\begin{equation}
\label{eq:ScalarFierz}
\textstyle{\frac 19}\big(2S+S^\flat)^2 =  
\textstyle{\frac 12}\big( J\!\cdot \! J-   H\!\cdot \! H\big)+ K\!:\!K^\top\,
\end{equation}
with  $J\!\cdot \!J = \eta^{\mu\nu}J_\mu J_\nu$\,,
$H\!\cdot \!H = \eta^{\mu\nu}H_\mu H_\nu$\,and
$K\!:\!K^\top = \eta^{\mu\nu}\eta^{\rho\sigma}K_{\mu\rho} K_{\sigma \nu}$\,. From these and similar identities, as discussed in the Appendix, it is possible to eliminate the tensor current $K_{\mu\nu}$, namely
\begin{equation}
K_{\mu\nu}= -\textstyle{\frac 13}(S-S^\flat)\eta_{\mu\nu} 
-\textstyle{\frac 34}\displaystyle{\frac{(J_\mu+H_\mu)(J_\nu-H_\nu)}{(S-S^\flat)}}\,.
\end{equation}
The algebraically independent currents are thus $S\,, S^\flat\,, J_\mu$\,, and $H_\mu$\,,
subject to the single constraint (either from the trace of $K_{\mu\nu}$\,, or by substitution for  $K\!:\!K^\top$ in the above scalar equation)\,,
\begin{equation}
\textstyle{\frac 14}\big( J\!\cdot\! J-   H\!\cdot\!H\big) + 
\textstyle{\frac 19}\big(S-S^\flat\big)\big(4S-S^\flat\big) =0\,,\label{Z and Z' Fierz constraint}
\end{equation}
which can itself be regarded as a condition to eliminate the scalar combination
$(4S-S^\flat)$ in terms of $(S-S^\flat)$, for example.
\section{Inversion of the DKP equation for $A_\mu$ and $F_{\mu\nu}$. }
\label{sec:DKPinv}
As mentioned in the introduction, the algebraic inversion of the DKP equation
proceeds by indirect algebraic manipulation rather than direct matrix inversion.
By pre-multiplying the DKP equation with chosen elements
$\Phibar \Delta \times \cdots$\, and combining these with the corresponding complex conjugate forms (given that $A_\mu$ is real), and the algebraic identities established above, the form of $A_\mu$ itself, and hence of the field strength $F_{\mu\nu}$\,, can be derived, as we now show.

Starting with the DKP equation and its complex conjugate,
\begin{align}
\big(\i \beta^\mu \partial_\mu -e \beta^\mu A_\mu -m\big)\Phi =&\,0\,,\\
\Phibar\big(\i \beta^\mu \overleftarrow{\partial_\mu} +e \beta^\mu A_\mu +m\big) =&\,0\,,
\end{align}
and pre-and post-multiplying by $\Phibar$\,, $\Phi$ and $\Phibar \beta^2$\,, $\beta^2 \Phi$\,, we obtain the two pairs of relations,
\begin{align}
&\partial_\mu J^\mu=0, &&eJ^\mu A_\mu=\textstyle{\frac 12}\i\big(\Phibar \beta^\mu(\partial_\mu\Phi)-(\partial_\mu \Phibar)\beta^\mu \Phi\big)-mS,\label{DKP J-constraints}\\
&\partial_\mu H^\mu=\textstyle{\frac 13}\i m\big(4S^\flat-10S\big), &&eH^\mu A_\mu=\textstyle{\frac 12}\i\big(\Phibar \bedt^\mu(\partial_\mu\Phi)-(\partial_\mu \Phibar)\bedt^\mu \Phi\big),\label{DKP H-constraints}
\end{align}
which entail the standard DKP current conservation condition, and also a companion current non-conservation condition,
as well as additional vector-gauge potential and companion vector-gauge potential quadratic constraints. Here the product relations in the DKP algebra
\begin{equation}
\beta^2 \beta_\mu = \textstyle{\frac 52} \beta_\mu - \textstyle{\frac 32} \bedt_\mu\,,
\quad \beta_\mu \beta^2 = \textstyle{\frac 52} \beta_\mu + \textstyle{\frac 32} \bedt_\mu\,,
\end{equation}
have been used (see Appendix).

Repeating this procedure, in this case by pre-and post-multiplication with 
$\Phibar\beta^\nu$\,, $\beta^\nu\Phi$ and $\Phibar \bedt{}^\nu$\,, $\bedt{}^\nu \Phi$\,,
leads similarly to two pairs of relations, expressing quadratic tensor current-gauge potential constraints 
on $e(K^{\mu\nu}\pm K^{\nu\mu})A_\nu$\,. In the second pair, however, the additional inhomogeneous
term in the relevant product relations,
\begin{equation}
\beta_\mu\bedt_\nu = -\beta_\mu\beta_\nu -\textstyle{\frac 23}\eta_{\mu\nu}({\mathbb I} - \beta^2) =-\bedt_\mu\beta_\nu\,,
\end{equation}
throws up a contribution proportional to $e A_\mu(S-S^\flat)$\,. Elimination of the $e(K^{\mu\nu}-K^{\nu\mu})A_\nu$\, tensor current contraction terms 
yields an equation for the companion vector as a gradient of the scalar current,
\begin{equation}
\label{eq:Helim}
H_\mu = \frac{\i}{3m}\partial_\mu(S-S^\flat)\,,
\end{equation}
while elimination of the $e(K^{\mu\nu}+K^{\nu\mu})A_\nu$\, tensor current contraction terms allows the gauge potential to be written as
\begin{equation}
A_\mu = \frac{3m}{2e}\frac{J_\mu}{(S-S^\flat)}+
\frac{1}{2e}\frac{\i (\Phibar(\partial_\mu\Phi) - (\partial_\mu \Phibar) \Phi)-\i (\Phibar \beta^2(\partial_\mu\Phi) - (\partial_\mu \Phibar) \beta^2 \Phi)}{(S -S^\flat)}\,.
\end{equation}
In this expression, the first term contains the gauge invariant, conserved current four-vector,
whereas the second, gauge-dependent, term contains derivatives acting `internally' on the DKP wavefunction itself, and so is not in bilinear form. 

The gauge dependence can still be 
accommodated in bilinear form, by introducing a further 15 complex bilinear currents associated with the corresponding symmetric DKP generators
$\eta\,, \eta\beta_\mu$\,, and $\eta \{\beta_\mu,\beta_\nu\}$\,. Defining
$\Phitld := \Phi^\top \eta$\,, these are $\Stld:=\Phitld \Phi $\,,
$\Jtld_\mu :=\Phitld \beta_\mu\Phi  $\,, and 
$\Ktld_{\mu\nu}:=\Phitld \beta_\mu\beta_\nu \Phi \equiv \Ktld_{\nu\mu} $\,
(with $\Htld_\mu :=\Phitld \bedt_\mu\Phi \equiv 0 $\,).

A Fierz-DKP rearrangement identity
for $\Phi\Phitld$ in terms of these complex currents, equivalent to that given above for $\Phi\Phibar$ in terms of hermitian currents, is derived in the Appendix from the general identity
for $\Phi\oPsi$ given there, by taking  $\Psi = \Phi^*$\,,
so that $\oPsi = \Psi^\dagger \eta = \Phi^\top \eta \equiv \Phitld$\,.
The coefficients are in fact identical in form (see equation (\ref{eq:FDKPidentityC})), with the omission of the $\bedt_\mu$ term. In view of the special form of the gauge dependent part of the expression for $A_\mu$, we require only a single special case, however: defining $\zeta:={\mathbb I}-\beta^2$ and
$\Ztld := \Stld-\Stld{}^\flat$\,, $Z:= S-S^\flat$\,,  we find using the $\beta$ identities
given in the Appendix,
\begin{equation}
\zeta \Phi\Phitld \zeta = \Ztld \zeta\,,
\end{equation}
which can be used to transcribe the expression into complex bilinear currents, as follows:
\begin{equation}
\frac{\Phibar \zeta\partial_\mu \Phi}{Z} =\frac{\Phibar \zeta\partial_\mu \Phi\cdot Z}{Z^2}\,,
\end{equation}
wherein (inserting the transpose of the second factor and rearranging  
using equation (\ref{eq:EtaTransposeEq}))
\begin{align}
Z^2 =&\, (\Phibar \zeta \Phi )(\Phibar \zeta \Phi ) = \Phibar \zeta(\Phi\Phitld) \zeta \Phi^* = (\Phibar \zeta \Phi^*) \Ztld \equiv \Ztld{}^*\Ztld \,.
\end{align}
Similarly
\begin{align}
\Phibar \zeta\partial_\mu \Phi\cdot Z =&\, (\Phibar \zeta (\partial_\mu\Phi \Phitld) \zeta \Phi^* ) = \textstyle{\frac 12}  (\Phibar \zeta \partial_\mu(\Phi \Phitld) \zeta \Phi^* ) =\textstyle{\frac 12} (\Phibar \zeta \Phi^*) \partial_\mu \Ztld\equiv \textstyle{\frac 12} \Ztld{}^* \partial_\mu \Ztld \,,
\end{align}
so that 
\begin{equation}
\frac{\i(\Phibar \zeta\partial_\mu \Phi- \partial_\mu \Phibar \zeta\Phi)}{Z}=
\frac 12\i\left(\frac{\partial_\mu \Ztld}{\Ztld}- \frac{\partial_\mu \Ztld{}^*}{\Ztld{}^*}\right)\,.
\end{equation}
Thus the additional gauge-dependent part in the expression for $A_\mu$ above 
can be written formally in terms of the imaginary part of $\partial_\mu(\ln \Ztld)$\,, and so is indeed a pure gauge which will not contribute to the field strength. Making the choice $\Ztld=\Ztld{}^*\equiv Z$, we have therefore in this gauge
\begin{equation}
\label{eq:InvPot}
A_\mu = \frac{3m}{2e}\frac{J_\mu}{(S-S^\flat)}\,.
\end{equation}
Using the equation (\ref{eq:Helim}) above for the companion vector current, the field strength $F_{\mu\nu} := \partial_\mu A_\nu -\partial_\nu A_\mu$\, becomes
\begin{equation}
\label{eq:InvFstr}
F_{\mu\nu} = \frac{3m}{2e}\frac{D_{[\mu} J_{\nu]}}{(S-S^\flat)}\,,\quad
\mbox{with}\quad D_\mu := \partial_\mu +3m\i\frac{H_\mu}{(S-S^\flat)}\,.
\end{equation}
Equations (\ref{eq:InvPot}) and (\ref{eq:InvFstr}) complete the task of algebraic inversion
of the self-interacting DKP equation, expressing the field quantities (the gauge potential (for a specific gauge choice) and its field strength) in terms of gauge invariant bilinear DKP currents and their derivatives. Given that the source term $e J$ for the equation of motion for the Abelian gauge field is as usual \cite{kemmer1939particle} given by the coupling to the DKP vector charge current, the field equations thereby also attain a gauge invariant form in the DKP bilinears. Using again
equation (\ref{eq:Helim})\,, the system reduces to a set of nonlinear differential equations in the vector current $J$ itself, together with the scalar density $Z:=(S-S^\flat)$\,. Introducing the locally scaled vector current ${\mathcal J}_\mu
:=J_\mu Z^{-1}$, so that $A_\mu =  (3m/2e) {\mathcal J}_\mu$\,, we have finally
\begin{align}
\big(\partial^2\delta_\mu{}^\nu-\partial_\mu\partial^\nu\big){\mathcal J}_\nu=\frac{2e^2}{m}Z{\mathcal J}_\mu\,.
\end{align}
Furthermore, from the LHS of (\ref{DKP J-constraints}) and (\ref{DKP H-constraints}), in addition with (\ref{Z and Z' Fierz constraint}) and (\ref{eq:Helim}), $Z$ and ${\mathcal J}^{\mu}$ must also satisfy the constraint equations:
\begin{equation}
Z\partial_{\mu}{\mathcal J}^{\mu}+{\mathcal J}^{\mu}\partial_{\mu}Z=0,
\end{equation}
\begin{equation}
{\mathcal J}_{\mu}{\mathcal J}^{\mu}=\frac{2}{9m^{2}}\left[\frac{\partial_{\mu}\partial^{\mu}Z}{Z}-\frac{(\partial_{\mu}Z)(\partial^{\mu}Z)}{2Z^{2}}\right]+\frac{4}{9}.
\end{equation}

\section{Conclusions}
\label{sec:Conc}
Since their original discovery, the DKP equations have remained candidate relativistic 
particle equations, and appear in traditional texts on quantum field theory 
\cite{akhiezerberestetskii1953quantum}
along with the Dirac equation, and the corresponding complex scalar Klein-Gordon, and massive vector Proca equations, with which they are usually regarded as equivalent (see for example \cite{green1953first} for a joint analysis of both cases)\,. While this is accepted at least in the free field case (for an historical review and 
further analysis see \cite{KrajcikNieto1977historical} and references therein),
it is an open question as to whether the interacting, second-quantized DKP theory,
including the 5 component case,
remains equivalent to standard field theories
(in curved spacetime for example \cite{Lunardi:2002op}).

In this paper we have given a gauge invariant reformulation of the self-coupled DKP equations, in terms of the set of real bilinear DKP currents. Our work provides the basis for a systematic examination of classical solutions of the self-interacting DKP equations under different spacetime symmetry group reductions \cite{Inglis2015}\,, and for the development of the bilinear method in an Einstein-Cartan setting \cite{Inglis_2019}\,. We expect analogous methods to be applicable also to the 10 component DKP system.
More generally, a functional change of variables would allow progress towards reformulation of the interacting DKP system as a nonlinear field theory. 
These topics will form the subject of future investigations.

\noindent
\textbf{Acknowledgements:}\\
The authors wish to thank Friedrich Hehl for discussions during the course of this work,
and Anthony Bracken for critical comments and drawing the authors' attention to reference \cite{harish1946correspondence}. PDJ acknowledges the kind hospitality of the theory group at K\"{o}ln during a visit.
%
%
%
%
%
%

\bibliographystyle{unsrt}

\appendix
\section{Algebraic structure in the 5 component DKP system} 
From the defining relations of the Kemmer algebra
\begin{equation}
\beta_\mu\beta_\rho \beta_\nu + \beta_\nu\beta_\rho \beta_\mu 
= \eta_{\mu\rho}\beta_\nu +\eta_{\nu\rho}\beta_\mu\,.
\end{equation}
and the trace identities, the following degree three and four relations follow by covariance:
\begin{align}\label{eq:CubicQuarticBetaId}
&\beta_{\lambda \mu\nu} = \, \textstyle{\frac 12}\big(\eta_{\lambda\mu}\beta_\nu + \eta_{\nu\mu}\beta_\lambda\big) + \textstyle{\frac 12}\big(\eta_{\nu\mu}\bedt_\lambda -\eta_{\lambda\mu}\bedt_\nu \big)\,, \\
&\beta_{ \kappa\lambda \mu\nu} = \,
\eta_{\lambda\mu}\beta_{ \kappa \nu} +{\textstyle{\frac 13}}
\big(\eta_{\kappa \lambda}\eta_{\mu\nu} - \eta_{\mu \lambda}\eta_{\kappa\nu} \big)\beta^2
-{\textstyle{\frac 13}}
\big(\eta_{\kappa \lambda}\eta_{\mu\nu} - \eta_{\mu \lambda}\eta_{\kappa\nu} \big){\mathbb I}\,,
\end{align}
where $\eta^{\mu\nu}\beta_\mu\beta_\nu := \beta^2$ ($\equiv \beta_0{}^2 -\beta_1{}^2-\beta_2{}^2-\beta_3{}^2$)\,, and
$\bedt_\mu :=\textstyle{\frac 13}\big(\beta_\mu \beta^2-\beta^2\beta_\mu\big)$\,.
Specific cases following from these basic identities are as follows:
\begin{align}
\bedt_\mu\bedt_\nu = &\, -\beta_\mu\beta_\nu\,; \\
\qquad \bedt_\mu \beta_\nu = &\,-\beta_\mu \bedt_\nu=\beta_\mu\beta_\nu - \textstyle{\frac 23}\eta_{\mu\nu}\big(\beta^2-{\mathbb I}\big)\,; \\
\qquad \beta_\mu\beta^2 = &\,\textstyle{\frac 52}\beta_\mu+\textstyle{\frac 32}\bedt_\mu\,,\\
\beta^2 \beta_\mu =&\, \textstyle{\frac 52}\beta_\mu-\textstyle{\frac 32}\bedt_\mu\; 
\end{align}
together with the contraction identities
\begin{align}
&\beta^\mu \beta_\rho \beta_\mu =\, \beta_\rho\,,\\
&\beta^\mu \beta_\rho \beta_\sigma \beta_\mu =\,\eta_{\rho \sigma}{\mathbb I}\,.
\end{align}
Further $\beta$ matrix identities at degree 5 and higher can be derived from (\ref{eq:CubicQuarticBetaId}) by associativity. In particular, defining the special
combination (scaled projection) $\zeta:={\mathbb I} - \beta^2$, we have
\begin{equation}
\zeta^2=-3\zeta,\quad \zeta\beta^2 \zeta = -12 \zeta\,,\quad
\zeta\beta_\mu\zeta =0\,, \quad 
\zeta\beta_\mu \beta_\nu \zeta = -3 \eta_{\mu\nu} \zeta\,.
\end{equation}

The basic Fierz-DKP rearrangement identity, equation (\ref{eq:FDKPidentity}), is derived as follows.
Any Hermitian combination $\Phi\overline{\Psi}$ of DKP wavefunctions $\Phi\,, \Psi$ may be expanded in terms of the basic set with arbitrary (real) coefficients
\begin{align}
\Phi \oPsi =&\, a {\mathbb I} + j^\mu \beta_\mu + \textstyle{\frac 12} k^{\mu\nu} \beta_\mu\beta_\nu
+ h^\mu\bedt_\mu\,,
\end{align}
and the coefficients determined by evaluating traces of the form 
$Tr\big(\Delta \Phi \oPsi \big)\equiv \oPsi \Delta \Phi$\, for and DKP matrix $\Delta$\,. For example, if $\Psi=\Phi$ and $\Delta = {\mathbb I}$\,, we have
trivially $\oPhi\Phi = 5a + k^\mu{}_\mu$\,, while with $\Delta = {\beta_\rho\beta_\sigma}$ we have $\oPhi\beta_\rho\beta_\sigma\Phi = 
2a \eta_{\rho\sigma} +   \textstyle{\frac 12}\big(k_{\sigma\rho}+\eta_{\rho\sigma}k^\mu{}_\mu\big)$\,. Solving these equations and the further 
relations following from tracing with $\beta_\rho$ and $\bedt_\sigma$ establishes the solution (\ref{eq:FDKPidentity}) (for the case $\Psi=\Phi$): 
\begin{align}
&a= \, \textstyle{\frac 59} \oPhi  \Phi - \textstyle{\frac 29}\oPhi \beta^2 \Phi\,;\\
&j_\mu =\, \textstyle{\frac 12}\oPhi\beta_{\mu}\Phi\,;\\
&h_\mu =\,- \textstyle{\frac 12}\oPhi\bedt_{\mu}\Phi\,;\\
&\textstyle{\frac 12}k_{\mu\nu}=\,
\oPhi \beta_{\nu\mu}\Phi -\eta_{\mu\nu}\big(\textstyle{\frac 29}\oPhi\Phi +\textstyle{\frac 19}\oPhi\beta^2\Phi\big)\,.
\end{align}
Homogeneous quadratic identities amongst the bilinear currents are derived in turn by
expanding the $\big(\Phi\oPhi \big)$ matrix in products of the form
$(\Phibar \Delta \Phi)\cdot (\Phibar \Delta' \Phi)$\,; for example 
\begin{align}
(\oPhi \Phi\big)^2 =&\,
\oPhi\Big( \big(\textstyle{\frac 59} S - \textstyle{\frac 29}S^\flat\big) {\mathbb I} 
+\textstyle{\frac 12}J^{\mu}  \beta_\mu + K^{\nu\mu} \beta_\mu\beta_\nu
 - \textstyle{\frac 12}H^{\mu}  \bedt_\mu 
 - \big(\textstyle{\frac 29}S +\textstyle{\frac 19}S^\flat\big) \beta^2          \Big)\Phi\,,
\end{align}
recovering the scalar identity equation (\ref{eq:ScalarFierz}) above.

For the Fierz-DKP rearrangement identity for complex bilinear currents required in the final 
reduced form of the inversion, the expansion basis is restricted to the symmetric
elements of the DKP algebra, $\eta\,, \eta\beta_\mu$\,, and $\eta \{\beta_\mu,\beta_\nu\}$\,. Taking the case $\Psi = \Phi^*$\,, in the expansion of $\Phi\oPsi$ (so that $\oPsi = \Psi^\dagger \eta = \Phi^\top \eta \equiv \Phitld$)\, the corresponding expression 
for $\Phi\Phitld$ in terms of these complex currents, equivalent to that given above for $\Phi\Phibar$ in terms of Hermitian currents, 
has coefficients  identical in form, with the omission of the $\bedt_\mu$ term ($\widetilde{h}{}^\mu=0$)\,, and such that
$\widetilde{k}{}^{\mu\nu}=\widetilde{k}{}^{\nu\mu}$\,:
\begin{align}
\label{eq:FDKPidentityC}
\Phi\Phitld = &\,
\big( \textstyle{\frac 59} \Stld - \textstyle{\frac 29}\Stld^\flat\big) {\mathbb I} 
+\textstyle{\frac 12}\Jtld^{\mu}  \beta_\mu + \Ktld^{\nu\mu} \beta_\mu\beta_\nu 
 - \big(\textstyle{\frac 29}\Stld +\textstyle{\frac 19}\Stld^\flat\big) \beta^2 \,.
\end{align}

\vfill
\end{document}